\documentstyle[12pt,psfig,html]{article}
\def\reference{\parskip 0pt\par\noindent\hangindent 0.5 truecm}
\def\uw{{\sc unswirf}}
\def\kms{km ${\rm s}^{-1}$}
\def\lesssim{\mathrel{\hbox{\rlap{\hbox{\lower4pt\hbox{$\sim$}}}\hbox{$<$}}}}
\def\gtrsim{\mathrel{\hbox{\rlap{\hbox{\lower4pt\hbox{$\sim$}}}\hbox{$>$}}}}
\def\farcm{\hbox{$.\mkern-4mu^\prime$}}
\def\farcs{\hbox{$.\!\!^{\prime\prime}$}}
\begin{document}
%
%
\title{UNSWIRF: A Tunable Imaging Spectrometer for the Near-Infrared}
%


\author{Stuart D. Ryder $^{1,2}$ \and
 Yin-Sheng Sun $^{1}$ \and
 Michael C. B. Ashley $^{1}$ \and
 Michael G. Burton $^{1}$ \and
 Lori E. Allen $^{1}$ \and
 John W. V. Storey $^{1}$
} 

\date{}
\maketitle

{\center
$^1$ School of Physics, University of New South Wales, Sydney 2052, Australia.\\
sun, mcba, mgb, lea, jwvs@newt.phys.unsw.edu.au\\[3mm]
$^2$ Current address: Joint Astronomy Centre, 660 N. A'Ohoku Place,
Hilo HI 96720, U.S.A.\\
sryder@jach.hawaii.edu\\[3mm]
}

%
\begin{abstract}
We describe the specifications, characteristics, calibration, and
analysis of data from the University of New South Wales Infrared
Fabry-Perot (\uw) etalon. \uw\ is a near-infrared tunable imaging
spectrometer, used primarily in conjunction with IRIS on the AAT, but
suitable for use as a visitor instrument at other telescopes. The
etalon delivers a resolving power in excess of 4000 (corresponding to
a velocity resolution $\sim75$~\kms), and allows imaging of fields up
to $100''$ in diameter on the AAT at any wavelength between 1.5 and
2.4~$\mu$m for which suitable blocking filters are available.
\end{abstract}


\bigskip

%
%

\section{Introduction}

The desire for high spectral resolution observations in the
near-infrared has been met with three main types of instrument.  The
traditional way of mapping an object at high spectral resolution is to
use a long-slit cooled grating spectrograph, and step the slit across
the sky. Although this technique records data at every spectral point
simultaneously, it is highly inefficient on extended objects if only a
single wavelength (or a small number of wavelengths) are of
interest. The Fourier Transform Spectrometer (FTS) works by Fourier
transforming an interferogram produced by a two-beam interferometer,
and can perform measurements over a large wavelength range (e.g., 0.9
-- 5.5~$\mu$m in the case of the CFHT FTS; Bohlender 1994).  The FTS
tends, however, to be mechanically large, complex, and expensive, and
is also not very efficient for monochromatic applications.
Furthermore, every pixel has the noise from the entire continuum in
it, and this noise is correlated from pixel to pixel. The Fabry-Perot
interferometer, by contrast, is small, has a high throughput (compared
to a grating of comparable size and resolving power; Jacquinot 1954),
and can deliver consistently high spectral resolution over a wide
field, which is imaged directly with an infrared array.

Fabry-Perot etalons have been successfully employed for narrow-band
imaging in the optical for many years (e.g., Atherton et al. 1982;
Bland \& Tully 1989; Jones \& Bland-Hawthorn 1997). However, it is
only the recent advent of low-noise, large-area detector arrays that
has made their use as tunable narrow-band filters for the
near-infrared particularly advantageous. There is clearly much to be
gained by observing emission lines in the infrared; for example, aside
from the reduction in extinction relative to the optical regime, many
rotational and vibrational transitions of molecules (such as H$_{2}$)
also become accessible.  Although many of the brighter Galactic
sources can be imaged using filters with fixed, narrow ($\Delta
\lambda / \lambda \sim 1$\%) bandpasses, the use of a Fabry-Perot
etalon with resolving power $\lambda / \Delta
\lambda \gtrsim 10^{3}$ confers a number of advantages, including

\begin{itemize}
\item the ability to resolve closely spaced lines, or resolve the
      line of interest from adjacent OH airglow or atmospheric absorption
      lines;
\item the ability to reveal velocity gradients, or even a complete velocity
      field, when Doppler motions exceed a few tens of \kms;
\item the reduced sky background and continuum flux passed to the detector.
      The resultant increase in the line-to-continuum ratio improves
      the measurement stability, and allows the sky to be sampled less often.
\end{itemize}

In this paper, we describe one such system, named \uw\ 
(University of New South Wales Infrared Fabry-Perot), which is
intended to complement the existing near-infrared imaging and
spectroscopic capabilities of IRIS\footnote{IRIS uses a
$128\times128$ HgCdTe array manufactured by Rockwell International
Science Centre, CA.} (Allen et~al. 1993) at the Anglo-Australian
Telescope (AAT), but which could also function as a visiting instrument at
other facilities (e.g., MSSSO 2.3~m, UKIRT). In the next section, we
give a brief overview of Fabry-Perot systems, and \uw\ in
particular. We then describe some of the novel approaches
taken to calibrate \uw\ and process the resultant data, and
give illustrations of some of the early scientific results obtained
with \uw.

\section{Overview of the Instrument}

\subsection{The Fabry-Perot Interferometer}

Thorough discussions of the principles underlying the Fabry-Perot
interferometer, and its use in astrophysics, can be found elsewhere
(e.g., Vaughan 1989; Bland \& Tully 1989), and only a few important
definitions will be given here. In essence, the Fabry-Perot
interferometer consists of a pair of identical transparent plates,
having plane-parallel internal faces of reflectivity $R$, separated by a
uniform spacing $d$. Peak transmission is attained over a series of
orders $n$ when

\begin{equation}
2 \mu d \cos \theta = n \lambda,  ~~~~~n = 0, 1, 2, 3, \ldots
\label{eq:peak}
\end{equation}

\noindent
where $\mu$ is the refractive index of the medium between the plates,
and $\theta$ is the angle relative to the normal of the incident beam
with wavelength $\lambda$. The spectral distance between two adjacent
orders $n$ and $n+1$ is called the Free Spectral Range (FSR), and is given by

\begin{equation}
\Delta \lambda_{FSR} = \frac{\lambda}{n} ~~.
\label{eq:fsr}
\end{equation}

\noindent
For a ``perfect'' Fabry-Perot etalon, the Full Width at Half-Maximum (FWHM)
of each order is

\begin{equation}
\Delta \lambda_{FWHM} = \frac{\lambda(1-R)}{n \pi R^{1/2}} ~~.
\end{equation}

\noindent
Thus the resolving power $\Re$ can be described as

\begin{equation}
\Re = \frac{\lambda}{\Delta \lambda_{FWHM}}
    = \frac{n \pi R^{1/2}}{1-R}
    = n F_{R}
\label{eq:rp}
\end{equation}

\noindent
where $F_{R}$ is called the reflection finesse. In practice, the true
``effective'' finesse $F_{e}$ of a Fabry-Perot system is always less
than the reflection finesse, due to surface defects in the coatings,
departures from plate parallelism, and the use of a converging, rather
than a parallel, incident beam. Replacing $F_{R}$ with $F_{e}$ in
equation~\ref{eq:rp} then allows the resolving power to be estimated
in the general case.

\subsection{The {\sc unswirf} etalon}

Table~\ref{t:fps} summarises the specifications of some currently
available near-IR imaging Fabry-Perot systems. As can be seen, various
combinations of resolving power, field of view, and tuning range
are available. The \uw\ etalon was specifically intended to meet the following
goals:
\begin{enumerate}
\item a high resolving power ($\Re \gtrsim 3000$); this was a trade-off
      between the amount of spectral scanning required at high resolution
      to fully sample the line, and the reduced line-to-continuum contrast
      and velocity information available at lower resolution;
\item a choice of pixel scales (e.g., $0\farcs77$~pixel$^{-1}$ or
      $0\farcs25$~pixel$^{-1}$, depending on the imaging optics selected
      within IRIS) matched to the seeing conditions and the clear aperture
      of the etalon;
\item the ability to cover lines in both the $H$~band (e.g., [Fe\,{\sc ii}]
      at 1.6440~$\mu$m) and in the $K$~band (e.g., H$_{2}~(1-0)~S(1)$ at
      2.1218~$\mu$m, H$_{2}~(2-1)~S(1)$ at 2.2477~$\mu$m, and Br$\gamma$
      at 2.1655~$\mu$m).
\end{enumerate}

\begin{table}
 \caption{Near-Infrared Imaging Fabry-Perot Systems$^{a}$}
 \label{t:fps}
 \begin{tabular}{llccccc}
\noalign{\medskip}
\hline
Instrument & Telescope & Band            & Resolving & Maximum &
Sensitivity$^{b}$ & Reference$^{c}$ \\
           &           &                 & Power $\Re$ & Field   &
                  &           \\
\hline
Cornell    & Various   & $K$             & 3300             & $18''\times19''$
& 7   & 1 \\
FAST       & Various   & $K$; $K$        & 1000; 2700       & $43''$ circle &
  5   & 2 \\
FINAC      & CRL 1.5~m & $J$; $K$; $K$   & 680; 1250; 12000 & $4'\times 4'$ &
 50   & 3 \\
IRAC       & ESO 2.2~m & $K$             & 1400             & $180''$ circle &
  5   & 4 \\
IRCAM3     & UKIRT     & $K$             &  860             & $73''\times 73''$
& 4  & 5 \\
NASM/NRL   & WIRO      & $J+H$; $K$      & 800; 800         & $64''\times64''$
& $<20$  & 6 \\
\hline
UNSWIRF & AAT          & $H+K$           & 4000             & $100''$ circle &
  5   & 7 \\
\hline
\end{tabular}
\begin{flushleft}
$^{a}$Characteristics of individual etalons used in each instrument are separated by semi-colons.\\
$^{b}3\sigma$ detection in 1000~s on-line integration in $K$ band, in units
of $10^{-16}$~ergs~cm$^{-2}$~s$^{-1}$~arcsec$^{-2}$, surmised from
the given reference.\\
$^{c}$1) Herbst et al. 1990; 2) Krabbe et al. 1993; 3) Sugai et al. 1994;
         4) Lidman et al. 1997; 5) Geballe 1997; 6) Satyapal et al. 1995;
         7) This paper.
\end{flushleft}
\end{table}

At the heart of \uw\ is a model ET-70WF etalon, manufactured by
Queensgate Instruments (UK) Ltd., with a clear aperture diameter of
70~mm. The plates are made from water-free fused silica, with a
matched surface quality of $\lambda / 200$ (for $\lambda = 633$~nm,
before coating). A series of multilayer dielectric coatings gives the
plates a reflectivity $R>97$\% all the way from 1.5 to $2.4~\mu$m.
The outer surface of each plate has a broad-band anti-reflection
coating applied.

As with most modern Fabry-Perot etalons, the separation and
parallelism of the plates is controlled to very high accuracy by
piezoelectric actuators, and servo-stabilised with capacitance
micrometers incorporated into the etalon itself. The Anglo-Australian
Observatory's Queensgate CS-100 servo-controller is capable of
maintaining the etalon spacing and parallelism to better than $\Delta
\lambda_{FSR} / 10^{4}$.  An IBM-compatible 286 PC rides in the
Cassegrain cage, along with an auxiliary electronics rack for
communication with both the CS-100 and an etalon translation
slide. Commands from the AAO MicroVAX~4000 computer to change the
etalon spacing as part of an observing sequence are relayed to the PC
by one of the AAO Sun workstations, and thence to the CS-100 via a
direct TTL logical level interface, with a typical response time
shorter than 1~ms.

A special mounting box has been constructed to go between the
Acquisition and Guide unit and IRIS at the Cassegrain focus of the
AAT. One side of this box holds a slide, controlled by a stepper
motor, which permits remote switching of the etalon in or out of the
beam with a positional accuracy of 1~$\mu$m. The other side of the
mounting box holds the polarimetry modules for IRISPOL (Hough et
al. 1994), making it possible for polarimetry to be performed in
conjunction with the Fabry-Perot if desired.  The etalon sits 140~mm
above the focal plane of the AAT, in an $f/36$ beam. This results in a
5\% reduction in the unvignetted field of view (compared to placement
in the focal plane), but no significant reduction in the spectral
resolution, owing to the small beam convergent angle (see also the
discussion in Greenhouse et al. 1997).

Besides making access to the etalon easier, the main benefit of
placing the etalon close to the focal plane (rather than close to the
pupil plane) is that each pixel ``sees'' only a very small part (just
12~mm$^{2}$) of the Fabry-Perot.  Operation in this
``pseudo-telecentric'' mode also results in a smaller change in
central wavelength across the field, as compared with operation in the
pupil plane. Any variations in plate spacing (i.e., departures from
flatness) translate into a variation in peak wavelength for that
region, rather than an overall decrease in finesse. Any such
variations in peak wavelength can be removed in the calibration
process. The main drawbacks of placing \uw\ outside the IRIS dewar
are the increased susceptibility to dust and to changes in the ambient
temperature and pressure, and a higher thermal background.

In direct-imaging mode, two optical configurations are available,
depending on the choice of re-imaging lens selected within IRIS
itself. The ``wide'' mode field of view is a circle $106''$ in
diameter, with $0\farcs77$ pixels, though the $128 \times 128$ pixel
array size of IRIS limits the usable field to just under $100''$.  In the
``intermediate'' mode, the pixel scale is $0\farcs25$~pixel$^{-1}$, and
the full $32'' \times 32''$ field is available. One limiting
factor on the capabilities of \uw\ is the availability of blocking
filters within IRIS. The standard narrow-band filters are
listed in Table~\ref{t:filters}. Since \uw\ is designed to work in
order $n\sim50$, equation~\ref{eq:fsr} shows that any of these filters
having bandwidths $\Delta \lambda_{FWHM} / \lambda < 0.02$ are adequate
for ensuring that only a single order is passed from the etalon to the
detector. Provided neither the continuum nor the night-sky emission is
too strong, the broader filters can still be used, though with a
corresponding reduction in signal-to-noise relative to a narrower
filter.

\begin{table}
 \caption{Blocking filters available in IRIS for use with \protect{\uw}}
 \label{t:filters}
\begin{center}
 \begin{tabular}{ccl}
\noalign{\medskip}
\hline
Central             &            Bandwidth   & Principal \\
Wavelength ($\mu$m) & ($\Delta \lambda_{FWHM} / \lambda$) & Line      \\
\hline
1.64  &  0.01  & [Fe\,{\sc ii}] (Galactic) \\
1.65  &  0.01  & [Fe\,{\sc ii}] ($0.002<z<0.006$) \\
1.74  &  0.01  & Br6 (n=10--4); H$_{2}~S(7)~1-0$ \\
2.12  &  0.01  & H$_{2}~S(1)~1-0$ \\
2.16  &  0.01  & Br$\gamma$ (n=7--4) \\
2.21  &  0.04  & Continuum \\
2.25  &  0.01  & H$_{2}~S(1)~2-1$ \\
2.34  &  0.04  & CO ${\rm v}=3-1$ and $4-2$ bands \\
\hline
\end{tabular}
\end{center}
\end{table}

Finally, it is also possible to insert the $H+K$ \'{e}chelle grating
and a slit in IRIS, and by scanning with \uw, build up a much higher
resolution spectrum of a source placed on the slit than would be
possible with the \'{e}chelle alone. Such a system could be used (with
or without the telescope), for example, to investigate the detailed
structure of the OH airglow emission spectrum.

\section{Observing with {\sc unswirf}\label{s:observe}}

\subsection{Wavelength Calibration}

The first step in commissioning \uw\ was to calibrate the relationship
between etalon spacing $d$ and peak transmitted wavelength $\lambda$.
This is complicated by a number of factors:
\begin{itemize}
\item the actual order $n$ being passed by the blocking filter is not
      necessarily known in advance;
\item the relationship between etalon spacing $d$, and the analog-to-digital
      units $Z$ ($-2048 < Z < +2047$) employed by the control computer and
      CS-100 must first be calibrated;
\item because of the wide tuning range demanded, and the multiple layer
      coatings involved, the effective depth where reflection
      occurs within the etalon coatings will change with wavelength, and
      so too will the apparent physical spacing of the plates. This manifests
      itself as a change in $\Delta \lambda_{FSR}$ with respect to an assumed
      value of the physical plate spacing (or equivalently, by a non-integral
      order number $n$). Let us define
      $\varepsilon_{\lambda}$ such that when the angle of incidence
      $\theta=0$, and the interplate medium is air at
      1~atmosphere pressure, equation~\ref{eq:peak} becomes
      \begin{equation}
      2(d+\varepsilon_{\lambda}) = n \lambda  ~~~~~n = 0, 1, 2, 3, \ldots
      \label{eq:epspeak}
      \end{equation}
      and thus
      \begin{equation}
      \Delta \lambda_{FSR} = \lambda_{i} - \lambda_{i+1}
                           = \frac{2(\varepsilon_{\lambda_{i}} - \varepsilon_
                              {\lambda_{i+1}}) + \lambda_{i+1}}{n_{i}}   .
      \label{eq:epsfsr}
      \end{equation}
      Thus, measurements of the FSR over the full range in wavelength and
      for multiple orders allows us to derive the order number $n$ and the
      wavelength variation of apparent plate separation $\varepsilon_
      {\lambda}$.
\end{itemize}
We now outline a novel calibration procedure that deals with these
issues, and replicates the actual observing configuration as closely as
possible, without the need for external calibration in the laboratory.

\begin{figure}
\begin{center}
\vspace{10cm}
\includegraphics{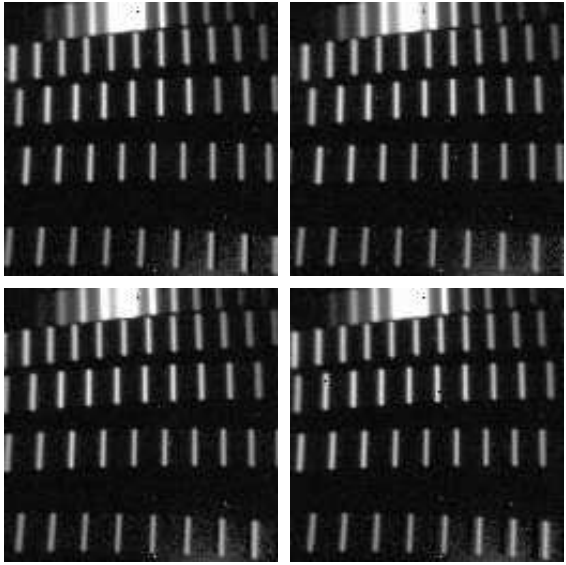}
\caption{The ``picket fence'' of orders produced using the $H+K$
\'{e}chelle and narrow slit of IRIS, a QI continuum lamp, and no
blocking filters with \uw\ is shown for etalon $Z=200$ ({\em top
left}), 400 ({\em top right}), 600 ({\em bottom left}), and 800 ({\em
bottom right}).  Note also how the spacing between adjacent orders of
the etalon ($\Delta \lambda_{FSR}$) increases with wavelength
(eqtn~\protect{\ref{eq:fsr}}). The wavelength ranges covered by each
complete \'{e}chelle order are (top to bottom, and from left to right)
1.44--1.70~$\mu$m, 1.62--1.91~$\mu$m, 1.86--2.18~$\mu$m, and
2.17--2.54~$\mu$m.}
\label{f:multiz}
\end{center}
\end{figure}

First, the etalon and all blocking filters are withdrawn from the
light path, and IRIS configured with the narrow ($1\farcs4$) slit and
its $H+K$ band \'{e}chelle grism.  Images of the emission-line spectra
produced by four separate lamps (Ar, Kr, Hg, and Xe) are then taken in
order to wavelength-calibrate the four complete \'{e}chelle orders
(covering the range $1.44-2.54~\mu$m after straightening) produced by
this grism. Next, the Fabry-Perot is inserted into the beam, the plate
spacing set to $Z=0$, and a continuum source (such as a quartz-iodine
lamp) used to illuminate the system.  Since there are no blocking
filters in place, all orders passed by the etalon are imaged,
resulting in the ``picket-fence'' appearance of
Figure~\ref{f:multiz}. Since wavelength as a function of position on
the array is already known, the position of each peak, and their
separations (i.e., $\Delta \lambda_{FSR}$) can be determined.  For
improved accuracy, these measurements are repeated with etalon
settings $Z=+1000, +900, +800, \ldots, -1000$, which causes the etalon
orders to shift position on the array, as illustrated in
Figure~\ref{f:multiz}.

\begin{figure}
\begin{center}
\vspace{7cm}
\includegraphics{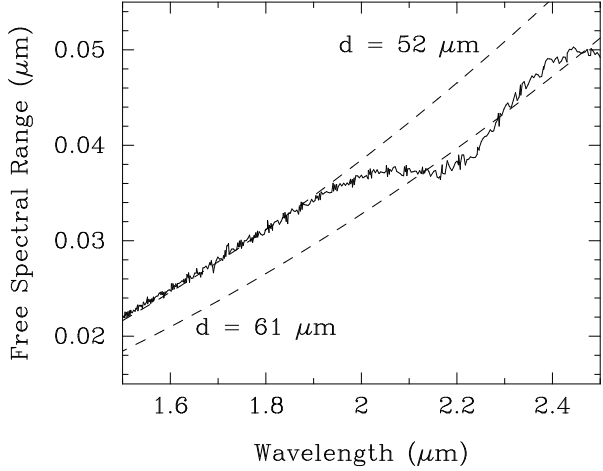}
\caption{Variation of Free Spectral Range $\Delta \lambda_{FSR}$ with
wavelength $\lambda$. The upper dashed line indicates a constant physical
etalon spacing of 52.0~$\mu$m, while the lower dashed line indicates a
constant physical etalon spacing of 61.0~$\mu$m.}
\label{f:fsrvl}
\end{center}
\end{figure}

As Figure~\ref{f:fsrvl} shows, the $\Delta\lambda_{FSR}$ of \uw\ is
indeed more complex than equation~\ref{eq:fsr} would suggest for a
simple dielectric coating. From equations~(\ref{eq:peak}) and (\ref{eq:fsr}),
it follows that for normal incidence in air

\begin{equation}
\Delta \lambda_{FSR} = \frac{\lambda^{2}}{2d} ~~.
\label{eq:dfsr}
\end{equation}

\noindent
The upper dashed line in Figure~\ref{f:fsrvl} corresponds to $\Delta
\lambda_{FSR}$ for a constant physical etalon spacing $d$ of 52.0~$\mu$m,
while the lower dashed line is for $d=61.0~\mu$m. For $\lambda<1.9~\mu$m,
the effective spacing between the plates ($d+\varepsilon_{\lambda}$) is
52.0~$\mu$m, but this grows rapidly at around 2~$\mu$m due to the
nature of the coatings, and is more like $61\pm2~\mu$m out to the long
wavelength cutoff of \uw.

The same data, analysed using equation~\ref{eq:epspeak}, indicate that
the change in spacing resulting from one step in $Z$ is

\begin{equation}
1~Z \equiv 0.9985~{\rm nm}
\end{equation}

Having determined $\varepsilon_{\lambda}$ over the full wavelength range
of \uw, the accuracy of this calibration has been tested by comparing
the predictions of equation~\ref{eq:epspeak} for the wavelengths of
the arc-lines measured earlier with their known wavelengths. We find
an r.m.s. accuracy of 0.04~nm, or $\sim8$\% of the instrumental
resolution.

\subsection{Parallelism, Resolving Power, and Finesse\label{s:prf}}

\begin{figure}
\begin{center}
\vspace{10cm}
\includegraphics{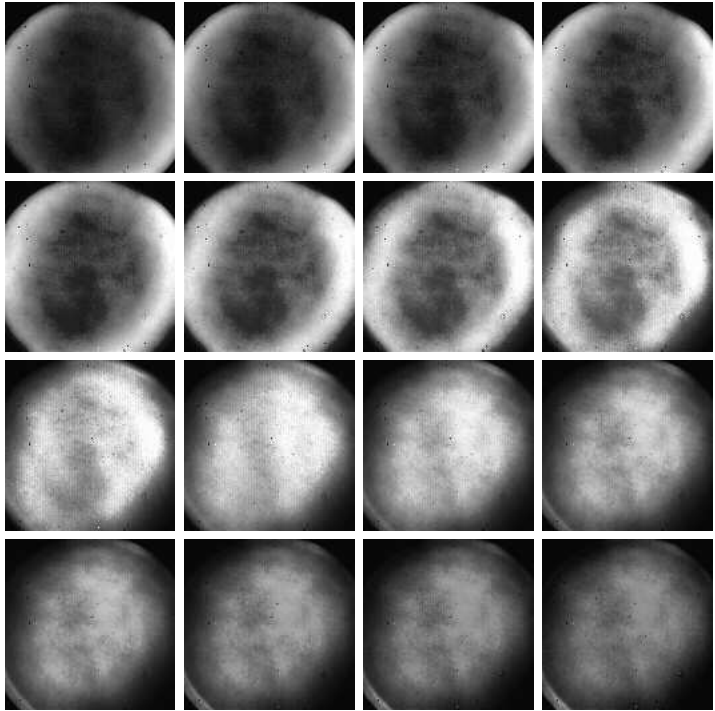}
\caption{UNSWIRF scan of the Ar 1.6520~$\mu$m line, incrementing the
etalon spacing by $2Z$ units each time. Etalon spacing increases from
left to right, and from top to bottom.}
\label{f:als}
\end{center}
\end{figure}

Having calibrated the relationship between $d$, $Z$, and $\lambda$, it
now becomes possible to set the etalon spacing for the peak
transmission of any required wavelength. The next step is to make the
plates as parallel as possible, and maintain this parallelism over the
full wavelength range, given the residual non-flatness of the plates
and/or their coatings. This can most easily be done by scanning in
wavelength across an unresolved emission line (e.g., from a discharge
lamp or from the OH airglow) and building up a ``cube'' of images,
with the third axis representing etalon spacing $Z$. For each spatial
pixel $(x,y)$ in the cube, the spectrum of intensity $I(Z)$ is
analysed with the {\sc unswfit} routine (Section~\ref{s:software}) to
determine the position $Z_{\rm peak}(x,y)$ and intensity of the
emission peak. Owing to the dependence on $\theta$ of the condition
for peak transmission (equation~\ref{eq:peak}), $Z_{\rm peak}(x,y)$
should be a maximum near the centre of the array, and decrease
towards the edges, as shown in Figure~\ref{f:als}.
From analysing such images, we find that the etalon surfaces are flat
to $\sim\lambda / 170$ at a wavelength of 1.65~$\mu$m, well within the
specification.

The way we have chosen to monitor the parallelism is to fit a tangent
plane to this surface, and then use the measured $x$- and $y$-slopes
to correct the parallelism settings passed to the etalon from the
CS-100. The parallelism has been found to be weakly, but repeatably,
dependent on $Z$ (and thus on $\lambda$, due to the fact that the
effective reflection at different wavelengths comes from different
depths within the coatings, and the coating thicknesses vary
slightly), and this is now accounted for by the observing control
software. After optimising the parallelism, we have calculated the
resolving power and effective finesse $F_{e}$ (equation~\ref{eq:rp})
of \uw\ using a series of discharge lamp lines over the available
wavelength range, as tabulated in Table~\ref{t:etpars}. The high
reflectivity $R$ of the plates will make $F_{R}$ the dominant
contributor to $F_{e}$. In addition, the throughput of the etalon at
each wavelength (except for 2.334~$\mu$m) has been measured by
comparing the peak intensities obtained with the etalon in and then
out of the beam.

\begin{table}
 \caption{Etalon parameters as a function of wavelength}
 \label{t:etpars}
\begin{center}
 \begin{tabular}{lcccrc}
\noalign{\medskip}
\hline
Line                & $n$ & $\Delta\lambda_{FWHM}$ & $\Re$ & $F_{e}$ &
Throughput \\
Wavelength ($\mu$m)$^{a}$ &     &      (nm)              &       &         &
    (\%)     \\
\hline
Ar 1.6437 & 64 & 0.378 & 4348 &  68 & 53 \\
Ar 1.6520 & 65 & 0.354 & 4666 &  72 & 62 \\
Kr 2.1165 & 51 & 0.386 & 5483 & 107 & 56 \\
Kr 2.1903 & 47 & 0.437 & 5011 & 107 & 70 \\
Ar 2.2077 & 50 & 0.470 & 4697 &  94 & 43 \\
Kr 2.2486 & 48 & 0.409 & 5497 & 115 & 43 \\
Ar 2.3133 & 47 & 0.560 & 4130 &  88 & 15 \\
Kr 2.3340 & 46 & 0.515 & 4532 &  99 & $\ldots$ \\
\hline
\end{tabular}
\begin{flushleft}
$^{a}$Line wavelengths in air.
\end{flushleft}
\end{center}
\end{table}

\begin{figure}
\begin{center}
\vspace{10cm}
\includegraphics{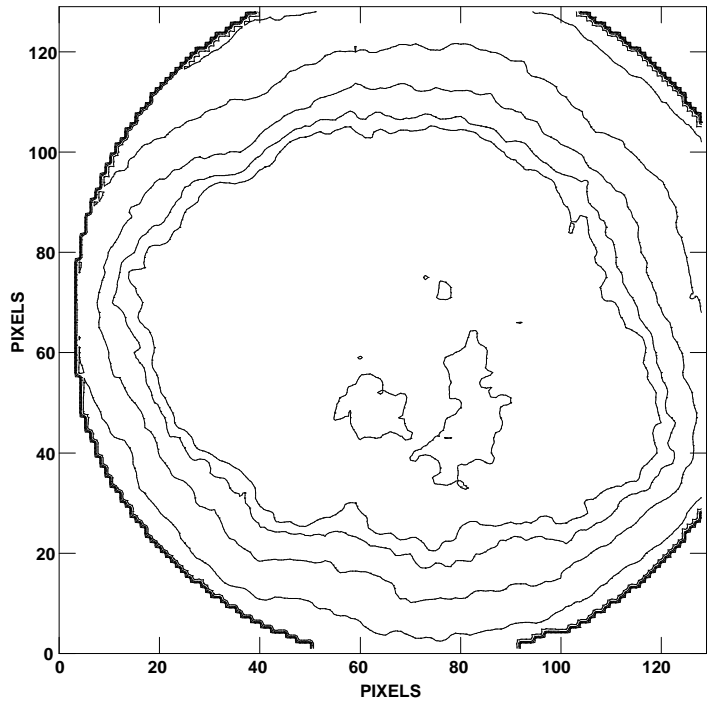}
\caption{Contour plot of the shift in transmitted wavelength with
position on the IRIS array for the Kr~2.1165~$\mu$m line, as a fraction
of the instrumental profile width $\Delta\lambda_{FWHM}$. The heavy black
line around the edge marks the unvignetted field of view of the etalon.
Beginning from the top right edge, the contours mark a shift of $-80$\%,
$-40$\%, $-20$\%, $-10$\%, and $-5$\% of $\Delta\lambda_{FWHM}$ relative
to the mean peak wavelength at the array centre.}
\label{f:uwshift}
\end{center}
\end{figure}

The shift in peak transmitted wavelength for the Kr~2.1165~$\mu$m
line, going from the centre to the edge of the etalon, is quite small
compared with the instrumental resolution.  As can be seen from
Figure~\ref{f:uwshift}, the shift is $<0.1 \Delta\lambda_{FWHM}$ over
the inner 90~pixel diameter, and still $<\Delta\lambda_{FWHM}$ over
the entire usable field of view. In fact, owing to possible
non-uniform illumination of the etalon by the discharge lamp,
Figure~\ref{f:uwshift} may slightly overestimate this shift. Thus,
\uw\ is virtually monochromatic, and can in principle be used as a
pure tunable line imaging filter, provided velocity gradients and
dispersions are small ($<50$~\kms), and the line centre wavelength is
known in advance. Otherwise, more extensive scanning in wavelength
will be necessary (but this of course furnishes, as a spinoff, the
velocity field). Although the ability of \uw\ to {\em resolve\/} lines
is limited by its instrumental profile width ($\sim60 - 70$~\kms,
depending on parallelism), we have found that the profile fitting
allows us to measure shifts in the position of the line peak
equivalent to velocity changes of $<10$~\kms, depending on
signal-to-noise of the data, profile shape, and plate parallelism.

\subsection{Observing strategies}


Observing with \uw\ is much like normal narrow-band imaging in the
near-infrared. As recommended by the IRIS manual (Allen 1993), readout
method 4 is employed, which breaks up each exposure into a series of
$\sim10$ Non-Destructive Readouts (NDRs), enabling on-the-fly bias
correction and linearisation, and yielding the lowest possible
read-noise (typically $\sim40$~e$^{-}$). To guarantee
background-limited performance, each exposure at a given etalon $Z$
setting is normally 120~s in duration at $K$ (180~s at $H$), broken up
into 12~NDRs. In order to reduce overheads, a complete scan in $Z$ on
the object is normally done before moving the telescope to an offset
sky position $\sim5'$ away, and repeating the sequence.  Except when
the scan crosses a strong, and rapidly varying, OH airglow line
(usually more of a problem in $H$-band than $K$-band), sky subtraction
is found to be quite adequate, even when sky frames are taken
$15-20$~minutes after the matching object frame.

An existing procedure for commanding telescope spatial offsets,
written in the AAO {\sc drama} environment (Bailey et al. 1995), has
been enhanced with the ability to request etalon spacing and
parallelism changes from the CS-100. To guard against possible drifts
in the \uw\ etalon parallelism (usually in response to changes in the
ambient temperature and/or humidity), the parallelism is normally
checked immediately prior to each night of observing, by scanning a
calibration lamp line close to the region of interest.

Because of the monochromatic nature of the infrared radiation reaching
the IRIS array from \uw, it is essential that matching sky exposures
and dome flatfields be obtained for {\em all\/} of the etalon $Z$ settings
used on an object, as otherwise severe fringing can result.
Similarly, bright spectroscopic standard stars need to be observed
once, and preferably twice at these same $Z$ settings. Although
accuracy of the photometry is usually limited by the sky and the array
to $\sim2$\% at best, it is necessary to determine the intensity
scaling of the continuum images relative to the line peak in order to
ensure proper continuum subtraction.

\subsection{Data Analysis\label{s:software}}

\begin{figure}
\begin{center}
\vspace{8cm}
\includegraphics{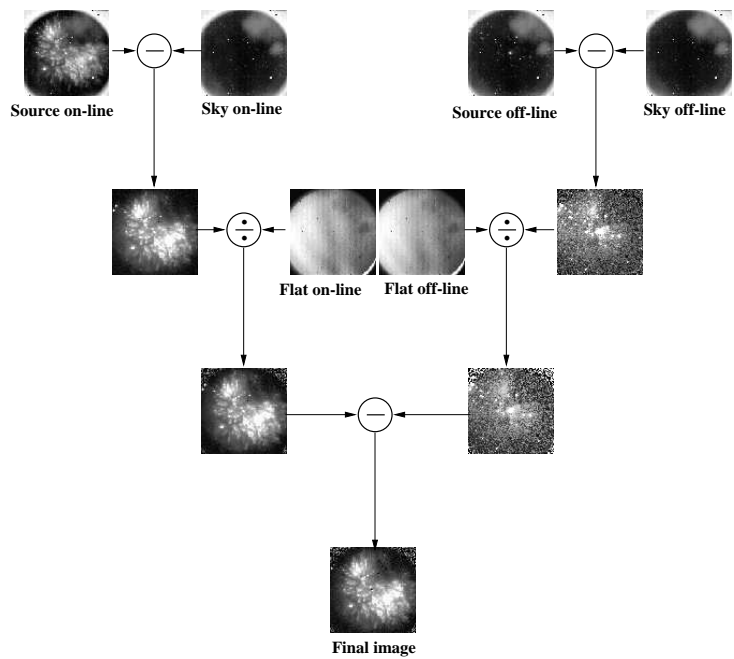}
\caption{Schematic diagram of data reduction steps, for monochromatic
imaging, using on- and off-line images of the H$_{2}~S(1)~1-0$ emission
around OMC-1.}
\label{f:drproc}
\end{center}
\end{figure}

A schematic of the basic data reduction procedure for \uw\ is shown in
Figure~\ref{f:drproc}, and begins with subtraction of a matching sky
frame, followed by division by a normalised matching dome
flat. Monochromatic imaging of simple sources then requires just a
scaling and subtraction of the off-line frame. For more complex
sources, a ``cube'' is constructed from a sequence of such
monochromatic images at a constant $Z$ interval, aligned to a common
spatial frame defined by field stars. The moments of this cube
(integrated line intensity $I(x,y)$, $Z$~position of the line peak
$Z_{\rm peak}(x,y)$, and the line width $\Delta\lambda_{FWHM}(x,y)$)
are extracted by fitting a Lorentzian to the spectrum at each spatial
pixel. In general, the line profile is usually too noisy to allow
three free parameters (the base level having already been set to
$\sim0$ by the off-line subtraction). Since in most cases the
emission-line profile will be unresolved by \uw, the line width
$\Delta\lambda_{FWHM}(x,y)$ can be assumed to be the same as the
instrumental profile width, as mapped by the calibration lamp line
scans, leaving only two free parameters in the fitting.  Finally, all
pixels in the vignetted corner regions of the moment maps, as well as
any pixels which fall below a specified intensity threshold, or for
which the fitted $Z_{\rm peak}(x,y)$ lies outside the actual $Z$ range
scanned (assuming that the observations did adequately span the line
of interest) are blanked out.

In order to streamline the processing of \uw\ data, a suite of
programs has been written using the IRAF\footnote{IRAF is distributed
by the National Optical Astronomy Observatories, which are operated by
the Association of Universities for Research in Astronomy, Inc., under
cooperative agreement with the National Science Foundation.}
environment. A listing of these programs and their functions is given
in Table~\ref{t:scripts}. With the exception of {\sc unswfit}, these
programs are scripts written in the IRAF Command Language (CL) which
execute a series of existing IRAF routines. The {\sc unswfit} task is
a purpose-written SPP (Subset Pre-Processor) program that uses a
Lorentzian-fitting algorithm supplied \mbox{by~F.~Valdes.}

\begin{table}
 \caption{Summary of \uw\ data reduction tasks}
 \label{t:scripts}
\begin{center}
 \begin{tabular}{ll}
\hline
Task name & \multicolumn{1}{c}{Purpose} \\
\hline
{\sc unswblank} & Execute {\sc unswfit}, then blank incongruous pixels in
                  output maps.\\
{\sc unswcal}   & Convert intensity from (e$^{-}$)~s$^{-1}$ to ergs~cm$^
                  {-2}$~s$^{-1}$~pixel$^{-1}$. \\
{\sc unswcube}  & Sky-subtract, flatfield, rotate, clean, align,
                  continuum-subtract\\
                & and stack a series of consecutive images into a datacube. \\
{\sc unswdisp}  & Display a ``movie'' of the datacube planes. \\
{\sc unswfit}   & Fit Lorentzian profiles to each datacube pixel, output maps                    of intensity, \\
                & wavelength shift, and profile width. \\
{\sc unswflats} & Produce and label flatfields. \\
{\sc unswlin}   & Convert data from ADUs to electrons. \\
{\sc unswmask}  & Mask a map using the same blanking as another map. \\
{\sc unswmerge} & Sort and stack a series of processed images into a datacube,
                  averaging \\
                & repeat data where available. \\
{\sc unswphot}  & Carry out aperture photometry on a sequence of standard star
                  images. \\
{\sc unswproc}  & Sky-subtract, flatfield, rotate, and clean a series of
                  consecutive images. \\
{\sc unswslope} & Execute {\sc unswfit}, fit tangent plane to wavelength shift
                  map, and compute \\
                & parallelism corrections. \\
{\sc unswspec}  & Plot a spectrum of intensity {\em vs} $Z$, averaged over a
                  range in $x$ and $y$.\\
{\sc unswvel}   & Correct wavelength shift map for instrumental shift, convert from $Z$ to \kms. \\
\hline
\end{tabular}
\end{center}
\end{table}

\section{Results}

\uw\ is a highly versatile facility, as illustrated by some of the
first science results achieved.  Since being commissioned
in 1996~February, it has been awarded a total of 35~nights in its first
3~semesters on the AAT. Among the projects currently underway (or
planned) are:

\begin{itemize}
\item Imaging and line-ratio mapping of supernova remnants, planetary nebulae,
      H\,{\sc ii}~regions, photodissociation regions (PDRs), and Herbig-Haro
      objects.
\item Photometry and dynamics of starburst and Seyfert galaxy nuclei.
\item Studies of extinction in star formation regions.
\item The search for redshifted UV/optical emission from primeval galaxies.
\end{itemize}

\begin{figure}
\begin{center}
\vspace{8cm}
\includegraphics{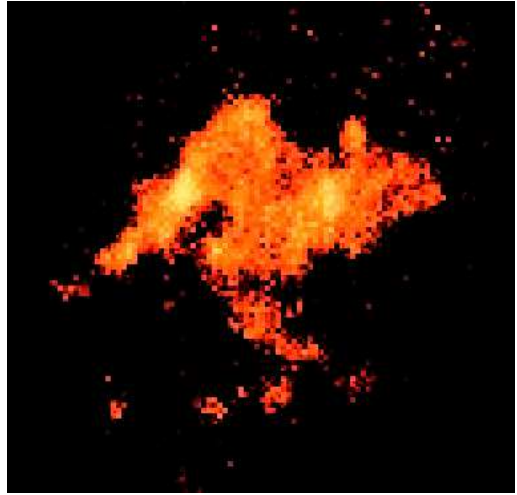}
\caption{Image of the H$_{2}~S(1)~1-0$ emission associated with a CO
outflow in Carina. The outline of this photodissociation region bears
an uncanny resemblance to one of the more abundant inhabitants of the
Warrumbungles National Park. The pixel scale is $0\farcs77$, and the image
spans $1\farcm5$. East is up, and North is to the right in this image.}
\label{f:kang}
\end{center}
\end{figure}

Figure~\ref{f:kang} was produced from some of the earliest data
obtained with \uw, and shows the emission from molecular hydrogen at
2.12~$\mu$m from a photodissociation region not far from the
``Keyhole'' Nebula  in Carina. Not surprisingly, this
region has earned the (unofficial) designation of the ``Kangaroo''
Nebula. This image was produced using \uw\ in its Line Imaging
Filter mode, by subtracting a continuum image from a single image
taken very near the line peak.

\begin{figure}
\begin{center}
\vspace{9cm}
\includegraphics{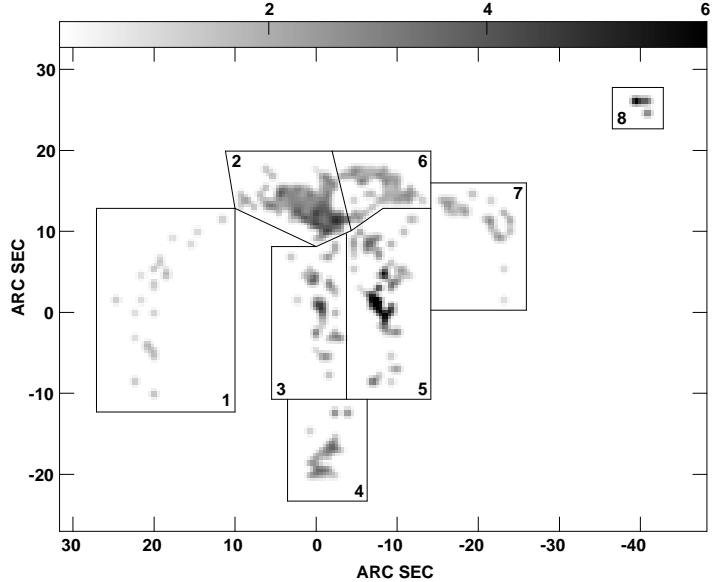}
\caption{Grey-scale map of the ratio of the H$_{2}~1-0~S(1)$ line at
2.12~$\mu$m to the H$_{2}~2-1~S(1)$ line at 2.25~$\mu$m in Parsamyan~18,
for all points in which a reliable detection (${\rm S/N}>3$) at 2.25~$\mu$m
was achieved. The coordinate system is relative to the position of a
$V=13.2$ B2-3e star, thought to be supplying the UV flux that pumps
much of the H$_{2}$~emission.}
\label{f:p18}
\end{center}
\end{figure}

\begin{figure}
\begin{center}
\vspace{6cm}
\includegraphics{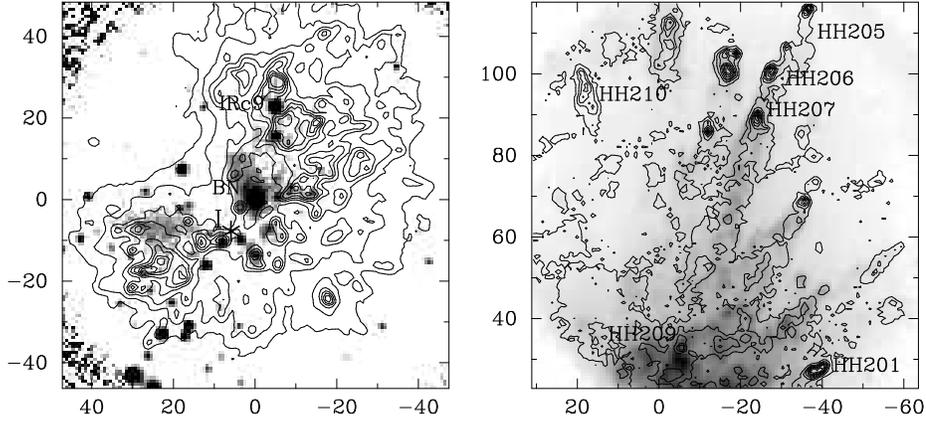}
\caption{Two images of the line emission from OMC-1. Offsets are in
arcseconds from the BN object. ({\em left}) H$_{2}~1-0~S(1)$ line (contours)
and adjacent
off-line continuum (grey-scale) in the core of OMC-1, showing the clumpy
nature of the line emission on arcsecond scales. ({\em right}) H$_{2}$ emission
to the NW of the core (grey-scale), overlaid with contours of [Fe\,{\sc ii}]
1.64~$\mu$m emission. Several of the [Fe\,{\sc ii}] emitting heads have been
identified with HH-object numbers. It can be seen that the fingers also
emit in [Fe\,{\sc ii}] as well as in H$_{2}$ (see Burton \& Stone (1998)
for a review of the H$_{2}$ emission from OMC-1).}
\label{f:orion}
\end{center}
\end{figure}

\uw\ is already helping to shed some light on the excitation mechanism
for H$_{2}$ in PDRs. Figure~\ref{f:p18} is a map of the
H$_{2}~S(1)~1-0$ / H$_{2}~S(1)~2-1$ intensity ratio in the reflection
nebula Parsamyan~18, obtained from scans of the 2.12 and 2.25~$\mu$m
lines with \uw\ (Ryder et~al.  1998). Values of the ratio $\sim3$ over
most of P~18 are indicative of UV-pumped fluorescence, while values
approaching 7 or more in the areas marked ``5'' and ``8'' are
consistent with an increased gas density and/or a contribution from
shocks.  The simultaneous velocity information provided by \uw\ has
allowed us to show that Region ``8'' is almost certainly excited by an
outflow source close to P~18, rather than being radiatively excited
like the other regions. Similar studies are also being carried out on
the ``elephant trunks'' of M16 (Allen et~al. 1998a), as well as the
``fingers'' emerging from the core of OMC-1 (Figure~\ref{f:orion};
Burton \& Stone 1998).

\begin{figure}
\begin{center}
\vspace{6cm}
\includegraphics{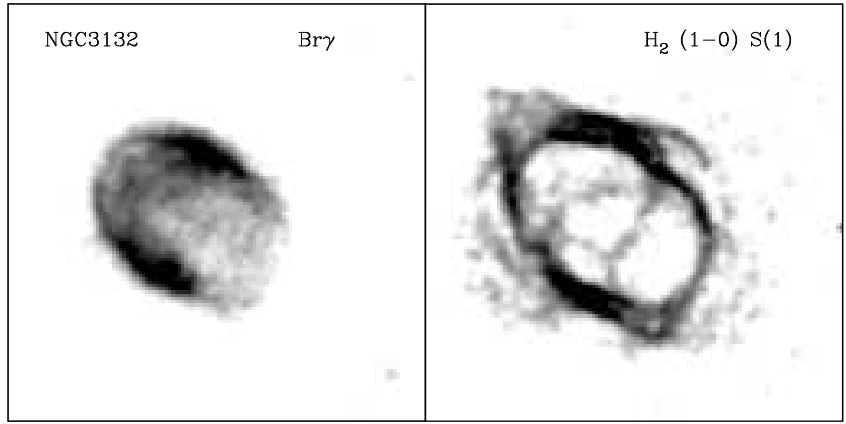}
\caption{Images of the Br$\gamma$ line emission from ionised hydrogen
({\em left}), and the H$_{2}$ 1--0 S(1) line emission from molecular
hydrogen ({\em right}) in the southern planetary nebula NGC~3132,
both imaged with \uw. Note how well the interface region between the
two regimes is defined, and the complex structure of the molecular emission.}
\label{f:n3132}
\end{center}
\end{figure}

One avenue of research to which \uw\ is particularly well-suited is
the excitation and dynamics of planetary nebulae, both young and
evolved.  Figure~\ref{f:n3132} compares the morphology of ionised and
warm molecular gas in NGC~3132 (Allen et al. 1998b). A similar study of
the H$_{2}$ emission in very low excitation (and therefore young)
planetary nebulae is also being conducted to complement an H$\alpha$
Snapshot survey with WFPC-2 on board the {\em HST} (Sahai \& Trauger
1996).

\begin{figure}
\begin{center}
\vspace{8cm}
\includegraphics{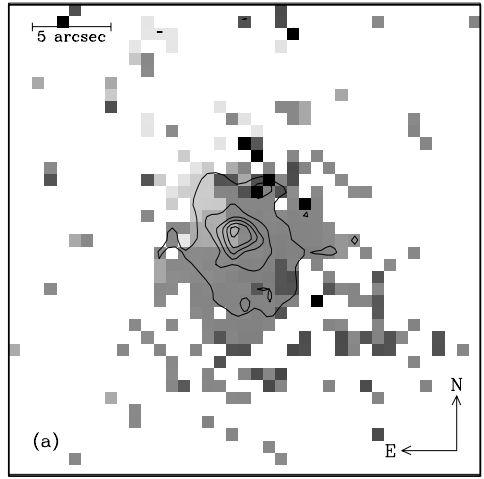}
\caption{Grey-scale velocity map for the central $30''$ of the Circinus
galaxy, derived from fitting of \uw\ scans to find the wavelength of the
H$_{2}$ 2.12~$\mu$m emission line peak. The contours indicate line intensity.}
\label{f:circvf}
\end{center}
\end{figure}

Another area of research where \uw\ is beginning to make inroads is in
the excitation and dynamics of active galactic nuclei and starburst
galaxies, normally heavily obscured by dust. Figure~\ref{f:circvf}
shows the inner velocity field, derived from \uw\ scans of the
H$_{2}$~2.12~$\mu$m line, in the Circinus galaxy, which is the closest
known Type~2 Seyfert galaxy. This has enabled the first direct measure
of the rotational velocity gradient near the nucleus of the Circinus
galaxy, and allows us to put an upper limit on the mass of the central
black hole of $1-2\times10^{7}$~M$_{\odot}$ (Davies et al. 1998).

\section{Conclusions}

We have successfully developed and commissioned the University of New
South Wales Infrared Fabry-Perot (\uw) on IRIS at the AAT.  Among
the notable characteristics of this system, which is based on a
Queensgate ET-70WF etalon, are its high Resolving Power ($\Re>4000$),
wide field (up to $100''$), and ability to be tuned across almost
the entire $H$ and $K$ bands. A novel method of {\em in-situ\/}
wavelength calibration has been applied, and a new suite of reduction
and analysis software in the {\sc iraf} environment has been
developed. Early results across a wide variety of sources have been
presented, and a number of diverse projects are now underway. Further
details are available from the \htmladdnormallinkfoot{\uw\ WWW page}
{http://www.phys.unsw.edu.au/$\sim$sdr/unswirf/UNSWIRF.html}.

\section*{Acknowledgements}

S.D.R. acknowledges the receipt of a UNSW Vice-Chancellor's
Postdoctoral Research Fellowship. \uw\ was funded by a grant from the
Australian Research Council. We are grateful to the Australian Time
Allocation Committee and the Director of the AAO for their generous
allocations of commissioning and observing time with \uw, and to
Stuart Lumsden and Tony Farrell for assistance with the observing software.
Chris Pietraszewski of Queensgate Instruments deserves special thanks
for his patient and enthusiastic support. Antonio Chrysostomou, Thomas
Geballe, and Colin Aspin provided useful comments on an earlier version
of this paper.

\section*{References}


\reference Allen, D. A. 1993, IRIS Users Guide (\htmladdnormallinkfoot{AAO
           User Manual 30a}
           {http://www.aao.gov.au/local/www/cgt/irisguide/iris\_guide.html})
\reference Allen, D. A., Barton, J. R., Burton, M. G., Davies, H., Farrell, T.,
           Gillingham, P., Lankshear, A., Lindner, P., Mayfield, D.,
           Meadows, V., Schafer, G., Shortridge, K., Spyromilio, J.,
           Straede, J., Waller, L., \& Whittard, D. 1993, PASA, 10, 298
\reference Allen, L. E., Burton, M. G., Ryder, S. D., Ashley, M. C. B.,
           \& Storey, J. W. V. 1998a, MNRAS, submitted
\reference Allen, L. E., Ashley, M. C. B., Ryder, S. D., Storey, J. W. V.,
           Sun, Y.-S., \& Burton, M. G. 1998b, in IAU Symposium 180: Planetary
           Nebulae (Dordrecht: Kluwer), in press
\reference Atherton, P. D., Taylor, K., Pike, C. D., Harmer, C. F. W.,
           Parker, N. M., \& Hook, R. N. 1982, MNRAS, 201, 661
\reference Bailey, J. A., Farrell, T. J., \& Shortridge, K. 1995, Proc. SPIE,
           2479, 62
\reference Bland, J., \& Tully, R. B. 1989, AJ, 98, 723
\reference Bohlender, D. A. 1994, Users' Manual for the CFHT Fourier
           Transform Spectrometer (\htmladdnormallinkfoot{Canada-France-Hawaii
           Telescope technical document}
           {http://www.cfht.hawaii.edu/manuals/fts/fts\_man.html})
\reference Burton, M. G., \& Stone, J. M. 1998, ASP Conf. Ser., in press
\reference Davies, R. I., Forbes, D. A., Ryder, S. D., Ashley, M. C. B.,
           Burton, M. G., Storey, J. W. V., Allen, L. E., Ward, M. J.,
           \& Norris, R. P. 1998, MNRAS, 293, 189
\reference Geballe, T. R. 1997, The 350~\kms\ Fabry-Perot Etalon for the
           $K$ Band (\htmladdnormallinkfoot{Joint Astronomy Centre technical
           document}{http://www.jach.hawaii.edu/$\sim$skl/fp.html})
\reference Greenhouse, M. A., Satyapal, S., Woodward, C. E., Fischer, J.,
           Thompson, K. L., Forrest, W. J., Pipher, J. L., Raines, N.,
           Smith, H. A., Watson, D. M., \& Rudy, R. J. 1997, ApJ, 476, 105
\reference Herbst, T. M., Graham, J. R., Beckwith, S., Tsutsui, K.,
           Soifer, B. T., \& Matthews, K. 1990, AJ, 99, 1773
\reference Hough, J. H., Chrysostomou, A., \& Bailey, J. A. 1994, ExA, 3, 127
\reference Jacquinot, P. 1954, J. Opt. Soc. Amer., 44, 761
\reference Jones, H., \& Bland-Hawthorn, J. 1997, PASA, 14, 8
\reference Krabbe, A., Rotaciuc, V., Storey, J. W. V., Cameron, M., Blietz, M.,
           Drapatz, S., Hofmann, R., S\"{a}mann, G., \& Genzel, R. 1993, PASP,
           105, 1472
\reference Lidman, C., Gredel, R., \& Moneti, A. 1997, IRAC2b User Manual V1.5
           (\htmladdnormallinkfoot{European Southern Observatory technical
            document}{http://www.ls.eso.org/lasilla/Telescopes/2p2T/E2p2M/IRAC2/irac2.html})
\reference Ryder, S. D., Allen, L. E., Burton, M. G., Ashley, M. C. B., \&
           Storey, J. W. V. 1998, MNRAS, 294, 338
\reference Sahai, R., \& Trauger, J. 1996, BAAS, 28, 1402
\reference Satyapal, S., Watson, D. M., Pipher, J. L., Forrest, W. J.,
           Coppenbarger, D., Raines, S. N., Libonate, S., Piche, F.,
           Greenhouse, M. A., Smith, H. A., Thompson, K. L., Fischer, J.,
           Woodward, C. E., \& Hodge, T. 1995, ApJ, 448, 611
\reference Sugai, H., Usuda, T., Kataza, H., Tanaka, M., Kawabata, H.,
           Inoue, M. Y., Takami, H., Aoki, T., \& Hiromoto, N. 1994, ApJ,
           427, 511
\reference Vaughan, J. M. 1989, The Fabry-Perot Interferometer (Philadelphia:
           Adam Hilger)
\reference Wall, L., Christiansen, T., \& Schwartz, R. L. 1996, Programming
           Perl, Second Edition (Sebastopol: O'Reilly \& Associates)

\end{document}